# Carrier-envelope phase stabilization of an Er:Yb:glass laser via feed-forward technique

**Randy Lemons,[1,2,*] Wei Liu,[1] Irene Fernandez de Fuentes,[1] Stefan Droste,[1] Günter Steinmeyer,[3] Charles G. Durfee,[2] Sergio Carbajo,[1,2]**

[1] *SLAC National Acceleratory Laboratory and Stanford University, 2575 Sand Hill Rd, Menlo Park, CA 94025, USA*
[2] *Colorado School of Mines, 1500 Illinois St, Golden, CO 80401, USA*
[3] *Max Born Institute for Nonlinear Optics and Short Pulse Spectroscopy, Max-Born-Straße 2a, 12489 Berlin, Germany*
**Corresponding author: rlemons@slac.stanford.edu*

**Few-cycle pulsed laser technology highlights the need for control and stabilization of the carrier-envelope phase (CEP) for applications requiring shot-to-shot timing and phase consistency. This general requirement has been achieved successfully in a number of free-space and fiber lasers via feedback and feed-forward methods. Expanding upon existing results, we demonstrate CEP stabilization through the feed-forward method applied to a SESAM mode-locked Er:Yb:glass laser at 1.55 µm with a measured ultralow timing jitter of 2.9 as (1 Hz – 3 MHz) and long-term stabilization over a duration of eight hours. Single-digit attosecond stabilization at telecom wavelengths opens a new direction in applications requiring ultra-stable frequency and time precision such as high-resolution spectroscopy and fiber timing networks.**

With the rise of few-cycle pulses in the femtosecond and attosecond regimes, stabilizing and controlling the carrier-envelope phase (CEP) has become increasingly important. Optical frequency metrology, for instance, requires optical frequency combs with well-known pulse characteristics [1,2]. In high harmonic generation and attosecond pulse generation, the intensity of the electric field is strongly coupled to the strength and shape of the generated light [3,4]. Optical frequency standards and clocks demonstrate the same challenges as optical frequency metrology to an even higher degree and greater control of the CEP will undoubtedly become important.

The shot-to-shot slippage of the CEP in mode-locked lasers largely arises from intracavity environmental conditions and optical power fluctuations. For a single shot, these conditions lead to a difference in the phase velocity and group velocity. This difference results in a phase shift ($\Delta\varphi$) of the carrier electric wave under the pulse envelope, often on the order of thousands of radians, which has been termed the group-phase offset. The relative shift of the envelope with respect to the peak of the electric field, constrained from zero to 2π, is the CEP. In free-running mode-locked lasers, the intracavity conditions are unstable and will cause the CEP to vary. The CEP results in a frequency shift of the mode-locked comb structure by the carrier-envelope offset (CEO) frequency [5], $f_{CEO}$, given by

$$f_{CEO} = \frac{\Delta\varphi \bmod 2\pi}{2\pi} f_{REP} \qquad (1)$$

where $f_{REP}$ is the repetition rate of the laser. By fixing $f_{CEO}$ to a known value, the phase slip from shot-to-shot can be stabilized.

Self-referencing techniques such as f-2f interferometry are most commonly used to detect $f_{CEO}$ [6–9]. With this technique, light is mixed according to

$$2(nf_{REP} + f_{CEO}) - 2nf_{REP} + f_{CEO} = f_{CEO} \qquad (2)$$

where $n$ is an integer referring to a single line in the optical comb, allowing access to the shot-to-shot value of the CEO frequency (beat signal). This signal can then be fed to various electronics for use control or stabilization [5].

The first subset of control techniques are *feedback* methods where modifications of pump power or cavity length are common [10,11]. These feedback methods serve to change the difference in the phase and group velocities. Pump power modulations change the actual velocities through intracavity nonlinearities while cavity adjustments simply adjust the path length and by extension the difference between the elements of the pulse traveling at the different velocities. Regardless of the feedback method, specialized electronics are necessary to maintain a lock between the measured CEO frequency and the driven changes.

Alternatively, the second subset of control techniques are *feed-forward (FF)* methods [12,13]. It is possible to stabilize and control the CEP of a free-running mode-locked laser not by acting internally on the cavity but on the output pulses. In this design, the beat signal is fed to an acousto-optic frequency shifter (AOFS) outside of the cavity to phase-modulate the laser frequency spectrum. In this way it is possible to adjust the comb directly and replace $f_{CEO}$ with a fixed frequency allowing for stabilization down the beam line, independent of cavity optics. Due to the direct use of the beat signal to drive the AOFS, only simple signal processing electronics are necessary to maintain stabilization. Short-term phase stabilization and long-term performance of the system are also decoupled due to the AOFS being an extra-cavity modification allowing intracavity conditions to evolve naturally. Additionally, the control bandwidth of the AOFS is on the order of several hundred kHz and is limited by the travel time of the acoustic wave from the transducer to the interaction zone inside the glass. However, this method is limited by small linear angular chirp introduced due to the acoustic grating and slow drifts of the beat signal away from optimal operation of the AOFS.

Both of these control methods have been used to varying effect; however, the mechanisms behind CEP-stabilization in mode-locked lasers are not entirely understood. Raabe et al [14] have highlighted some of the technical aspects of fiber laser systems that may lead to to excess phase-noise arising primarily from the mode-locking mechanism and not from the gain medium. In the case of semiconductor saturable absorber mirror (SESAM) mode-locked lasers, significant phase changes can occur from spurious intra-cavity emission noise due to its strong coupling to ultra-small variations in saturable absorber reflectivity. This effect is even sensitive enough to have been exploited as a feedback technique [15]. However, limited literature exists for Er lasers, regardless of the mode-locking mechanism, being stabilized via the FF technique. The lowest CEP jitter in a FF configuration to date is reported by Kundermann et al [16] at 120 mrad of integrated phase noise from 0.01 Hz to 1 MHz. However, we suspect that the noise improvement was hampered by the lack of amplification in the experimental design. In this manuscript, we intend to add onto existing literature [16,17] by presenting ultralow CEP jitter from an Er-doped mode-locked laser using the FF technique as well as improvements in the $f_{CEO}$ detection configuration.

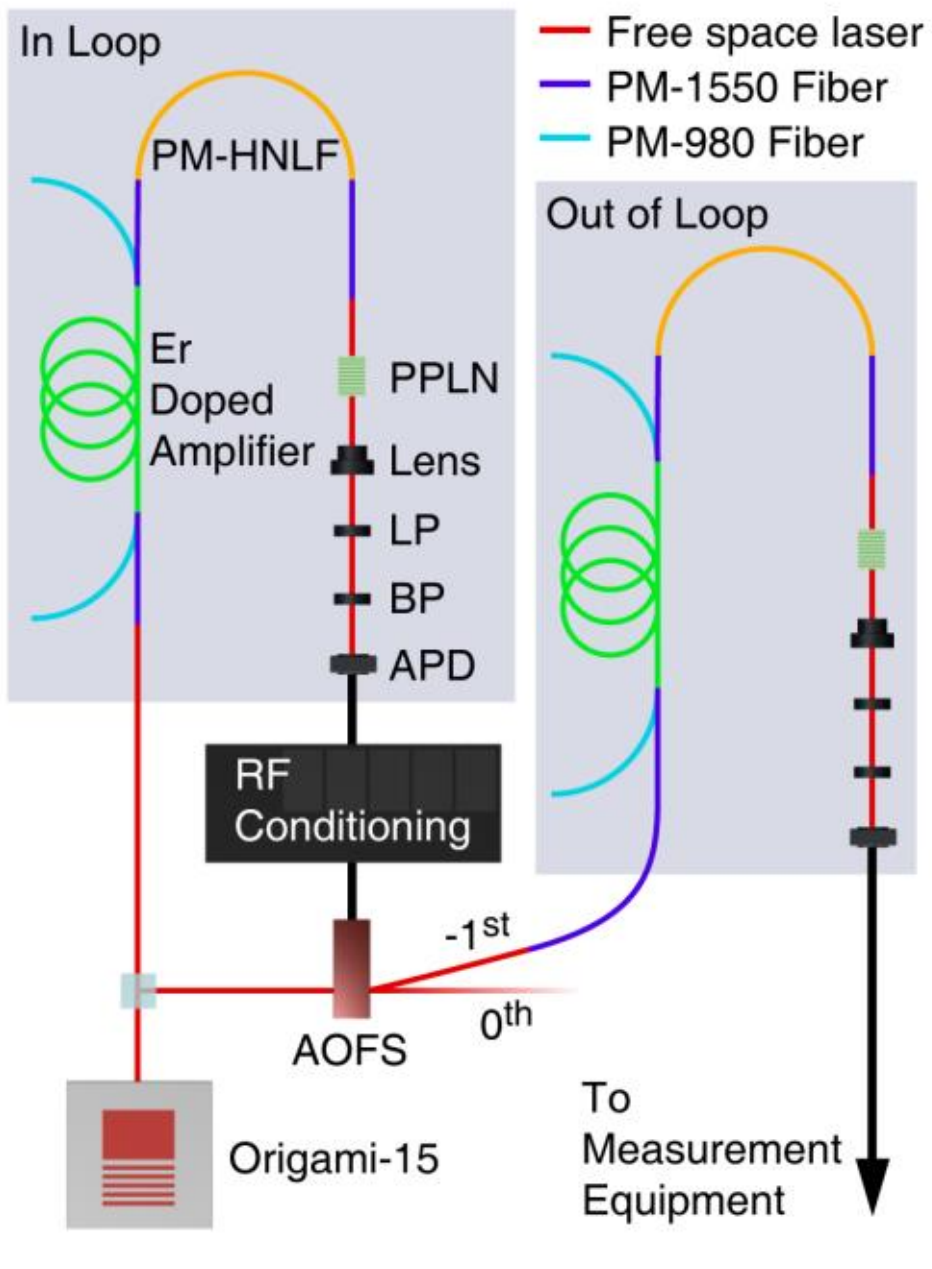

Fig. 1. Experimental setup with in-loop and out-of-loop beamlines. The Erbium doped amplifiers are fed with 14 pin butterfly diodes at 980 nm. Half waveplates (not pictured) are used to align the polarization to the fiber axes and the periodically poled lithium niobate (PPLN) crystal. LP: low pass filter at 1500 nm. BP: Bandpass filter centered at 1024 nm. PM-HNLF: Polarization maintaining highly nonlinear fiber.

For our investigation, we use a SESAM soliton mode-locked Er:Yb:glass laser oscillator (OneFive ORIGAMI-15) in our FF system (Fig. 1). The oscillator delivers 140 mW of power in 175 fs pulses at a repetition rate of 204 MHz with a spectral bandwidth of 14.9 nm centered around 1.55 μm.

Er:Yb:glass lasers serve as an excellent base for a CEP stabilization system due to the intrinsically low timing jitter [18]. This increased performance may be due to cooperative energy transfer, reduced back-conversion transfer, and up-conversion losses present in Er, Yb co-doped solid-state lasers [19]. In addition, the high power with short pulses intrinsic to the soliton mode-locked design [20] aids in large signal to noise ratio (SNR) inside the *f-2f* interferometers.

The light from the oscillator is split into two beamlines, one towards the in-loop (IL) feedback measurement, and the other through the AOFS (AA Opto-Electronic MGAS80-A1) and towards the out-of-loop (OOL) measurement. This AOFS is estimated to have a bandwidth of 500 kHz based on a manufacturer listed sound velocity of 2520 m/s. Both beamlines are coupled into stretcher fibers which is spliced to Er:fiber amplifiers. Amplification of the signal has been shown to increase performance by increasing the signal to noise ratio during measurement. This comes about due to a reduction in shot noise from an average increase in the number of available photons for interference and the resulting beat signal [21]. Our use of amplification before the *f-2f* allows use of a monolithic design, albeit at a reduction in the total number of interfering photons.

After amplification the pulses are recompressed in PM-1550 fiber to 60 fs with 250 mW of average power. High pulse power is necessary to achieve the required octave spanning spectrum via self-phase modulation in the HNLF. Spanning from 1000 nm to upwards of 2080 nm is possible with 250 mW out of the amplifiers. The spectrally broadened pulses are then coupled out to free space and frequency-doubled in magnesium-doped PPLNs tuned for second harmonic generation at 1024 nm. The light is then passed through optical band pass filters centered at 1024 nm and focused on to avalanche photo diodes (APD).

Temporal and spatial overlap between the frequencies at 1024 nm is necessary for detection of the heterodyne beat signal. Spatial overlap is ensured through the common path architecture. However, because all light is passed through the same fiber the phase velocity is not uniform at all frequencies. To address this, the optical power in the HNLF is chosen so that 1024 nm and 2048 nm are among the last frequencies generated. The temporal walk off is therefore small allowing a band on either side to be used in the *f-2f* scheme without needing corrections to timing differences.

The raw signal from the APD must be conditioned because the AOFS has an operational frequency of 80 ± 2.5 MHz and $f_{CEO}$ is not guaranteed to be in this range. The signal is first filtered to remove $f_{REP}$ and mixing products with $f_{CEO}$. The isolated $f_{CEO}$ with 40 dB SNR (RBW: 100 kHz) is then mixed with a local oscillator (LO) derived from a 10 MHz rubidium standard (Stanford Research Systems PRS10). The 10 MHz signal is passed into a divider to create a comb of frequencies increasing from 1.4 MHz in steps of 1.465 MHz. Depending on where $f_{CEO}$ rests, a different line of this comb is chosen for mixing with a tunable band pass filter. The final signal is therefore given by $f_{AOFS} = f_{CEO} + f_{LO} = 80$ MHz.

This mixed signal near 80 MHz is then amplified to 26 dBm to drive the AOFS in the OOL beam. The AOFS subtracts the drive signal from the frequency comb according to Eq. ***(3)*** and the power is shifted to the AOFS -1st diffraction order. The linear angular chirp that is introduced in the AOFS is mitigated by coupling into fiber 10 cm after the AOFS though it has been demonstrated that a properly designed prism can be used to compensate for chirp [13]. We operate the AOFS with an 80% diffraction efficiency.

$$\begin{aligned} \boldsymbol{f_{OOL}} &= \boldsymbol{f_n - f_{AOFS}} \\ &= (\boldsymbol{nf_{REP} + f_{CEO}}) - (\boldsymbol{f_{CEO} + f_{LO}}) \\ &= \boldsymbol{nf_{REP} - f_{LO}}, \end{aligned} \qquad \textbf{(3)}$$

In the OOL interferometer the heterodyne beat signal is located at $f_{LO}$ and the stability of the phase of the pulses is dependent on the stability of $f_{LO}$. The raw signal from OOL APD contains $f_{REP}$, $f_{LO}$, and the mixing products (Fig. 2). This signal is sent through a low pass filter to remove the strong signal at $f_{REP}$ for analysis.

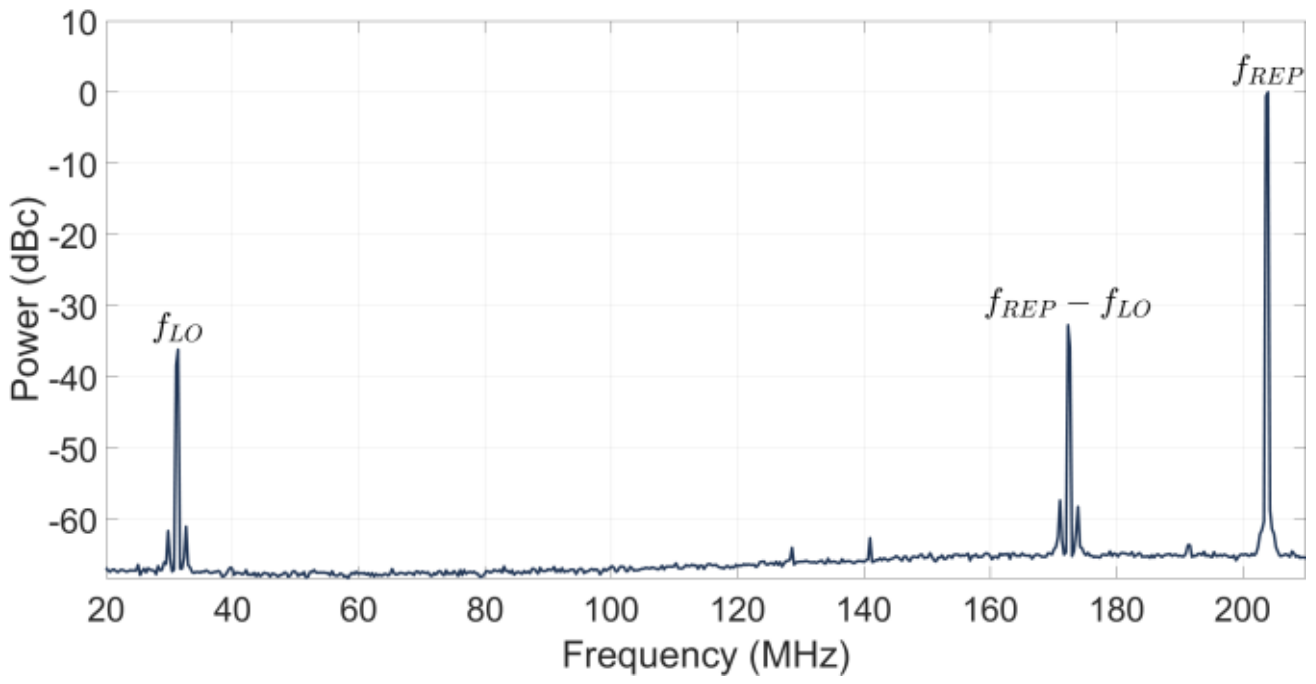

Fig. 2. RF spectrum (RBW: 100 kHz) of the stabilized $f_{OOL}$ before low-pass filtering to remove $f_{REP}$ and mixing products. $f_{LO}$ is seen here with 31 dB SNR with side bands from the generated reference comb.

The short-term phase noise measurements shown in Fig. 3 were made with a signal source analyzer (Rohde & Schwarz FSUP 26) with a low phase noise cross-correlation (XCORR) module. In addition, the 10 MHz clock signal was used as an external reference to eliminate any noise fluctuations due to drifts of the analyzers' LO. Each measurement is expressed as a single side band phase noise spectral density, $\mathcal{L}(f)$, in dBc made within the phase lock loop (PLL) mode. Locking bandwidth and carrier frequency search were set by the device. The integrated phase noise (IPN) in radians can be recovered from $\mathcal{L}(f)$ according to

$$\boldsymbol{IPN} = \sqrt{\boldsymbol{2}\int_{\boldsymbol{f_{low}}}^{\boldsymbol{f_{high}}} \boldsymbol{10^{\mathcal{L}(f)/10}}} \qquad \textbf{(4)}$$

where $f_{high}$ and $f_{low}$ represent the bounds of the measurement. Care must also be taken to integrate $\mathcal{L}(f)$ on a linear, rather than log scale necessitating the conversion within the integrand. Additionally, the IPN is conventionally defined as the integral of the spectral density of phase fluctuations [22], $S_{\varphi}(f)$, which is twice $\mathcal{L}(f)$ resulting in the leading factor of two [23].

Due to $10^4$ Hz level drifts in the free-running laser, it was not possible to collect a phase noise density or IPN measurement. The R&S FSUP 26 has a maximum PLL bandwidth for the noise measurement of 30 kHz therefore the free-running laser drift was at minimum larger than this value. The stabilized OOL signal in Fig. 3 displays an IPND of 25 mrad from 0.1 Hz to 3 MHz. This results in a rms phase jitter, $\Delta\varphi_{rms}$, of 20 as at 1.55 µm. However, the measurement seems to be dominated by $1/f$ spurious noise at integration times below 1 Hz. This is seen in all OOL characterization systems and can be attributed to quantum noise in the laser cavity [24]. Integrating instead from 1 Hz to 3 MHz, in order to evaluate the performance of the system apart from these drifts, leads to an IPND of 3.5 mrad and a $\Delta\varphi_{rms}$ of 2.9 as. Additionally, $f_{LO}$ displays a small increase in phase noise around 10 kHz that is absent in the OOL signal even though the noise floor above 1 kHz seems to be shared between the two signals. One possibility is that this small signal is not strong enough to be properly transmitted via the AOFS to the OOL light.

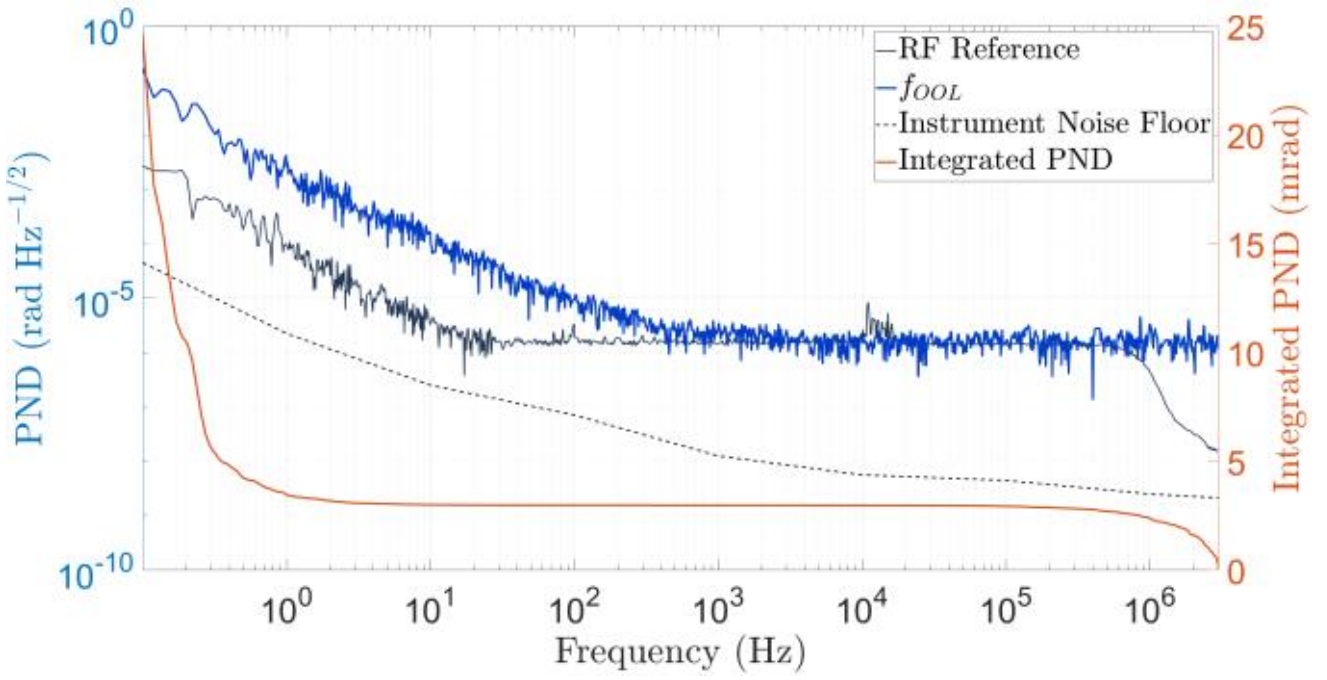

Fig. 3. Single side band phase noise spectral density (PND) plots of $f_{LO}$ (grey) and $f_{OOL}$ (black). IPN (orange) is shown on the right vertical axis.

Because of the natural slow drift from the beat signal, long-term stability of our system is currently maintained through adjustment of the pump power in the oscillator about every half hour to keep $f_{AOFS}$ within 80 ± 0.25 MHz. This perturbative manual adjustment introduces large amounts of phase noise. Additionally, as $f_{AOFS}$ drifts away from 80 MHz there is an increase in measured phase jitter. This will be readily resolved with a slow feedback PID controller on the pump power to keep $f_{AOFS}$ centered on 80 MHz in future work. Due to the millisecond time scale of the upper state lifetime of Er:Yb:glass, this should allow for loop bandwidths around the kHz level.

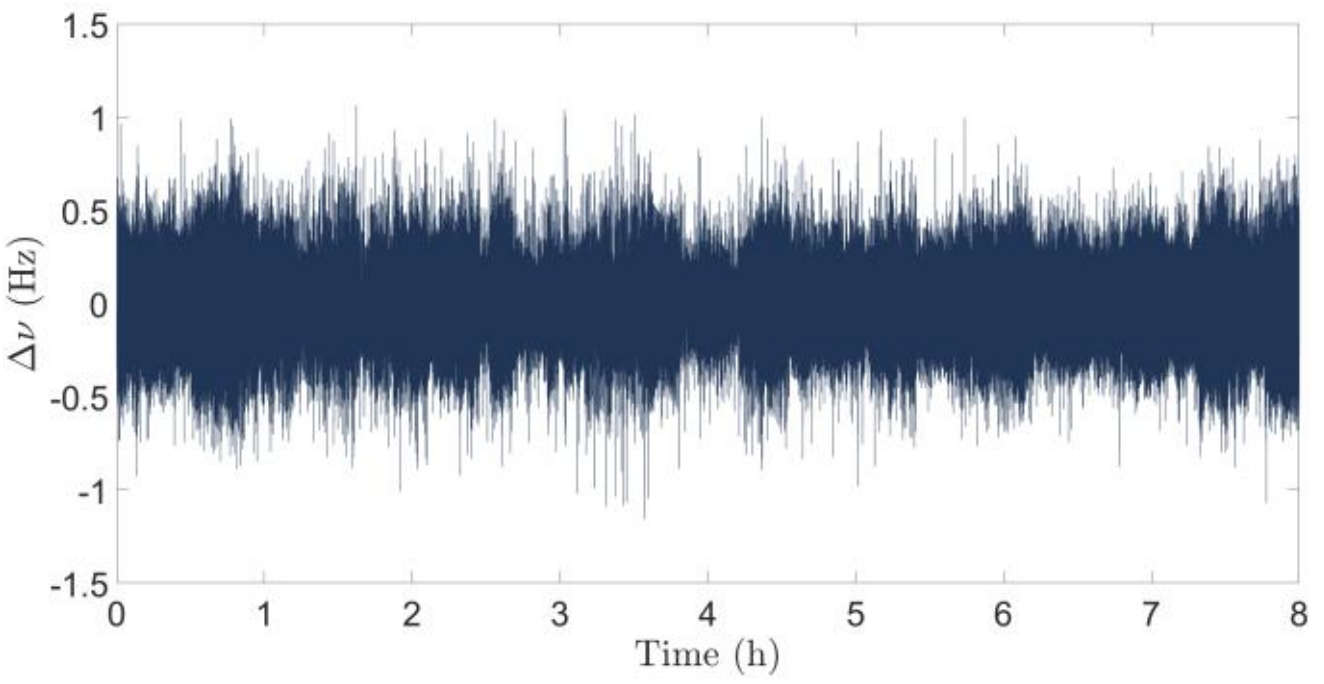

Fig. 4. Eight-hour frequency drift of the stabilized OOL signal relative to the set point of $f_{LO}$ showing 0.16 Hz rms variance during a typical work day.

The long-term performance can be quantified by mixing $f_{OOL}$ and $f_{LO}$ to isolate the phase jitter of the CEP from the phase slip of the LO. The deviation from zero is then a direct measure of the jitter of the CEP. Currently $f_{OOL}$ is measured as an indication of long-term stabilization, not phase performance. Fig. 4 demonstrates the continuous measurement of $f_{OOL}$.

In conclusion we have demonstrated CEP stabilization of an SESAM mode-locked Er:Yb:glass laser system through use of the FF f-2f self-referencing method. The reported IPN at 3.5 mrad (1 Hz - 3 MHz) is a significant improvement to previously reported values for Er lasers via the FF method. We associate our degree of stabilization to the intrinsically low timing jitter of Er:Yb:glass lasers in addition to the high SNR that is achieved in the IL f-2f interferometer. We expect that by addressing the drift noise below 1 Hz that full-band noise can be reduced and brought into agreement with the narrowed band.

**Funding.** This work was supported in part by the U.S. Department of Energy, Laboratory Directed Research and Development program at SLAC National Accelerator Laboratory, under Contract No. DE-AC02-76SF00515; This material is based upon work supported by the U.S. Department of Energy, Office of Science, Office of Workforce Development for Teachers and Scientists, Office of Science Graduate Student Research (SCGSR) program. The SCGSR program is administered by the Oak Ridge Institute for Science and Education (ORISE) for the DOE. ORISE is managed by ORAU under contract number DE-SC0014664. All opinions expressed in this paper are the author's and do not necessarily reflect the policies and views of DOE, ORAU, and ORISE.